\pdfoutput=1
\documentclass[aps,prl,floatfix,twocolumn,superscriptaddress]{revtex4-2}

\usepackage{physics}
\usepackage{amsfonts}
\usepackage{mathrsfs}
\usepackage{amsmath}
\usepackage{color}
\usepackage{graphicx}
\usepackage{bm}
\usepackage{amssymb}
\usepackage{upgreek}
\usepackage{xspace}
\usepackage{epstopdf}
\usepackage{dcolumn}
\usepackage{longtable}
\usepackage[colorlinks=true, pdfstartview=FitV, linkcolor=blue, citecolor=blue, urlcolor=blue]{hyperref}
\usepackage{hyperref}
\usepackage{comment}

\newcommand{\x}[1] { \mathrm{#1} }
\newcommand{\lch} { L_\x{ch} }
\newcommand{\rc} { R_\x{c} }
\newcommand{\lmfp} { l_\x{mfp} }
\newcommand{\gtyp} { G_\x{typ} }

\begin{document}
\title{Semiconductor Meta-Graphene and Valleytronics}

\author{Praveen Pai}
\affiliation{Department of Physics, University of Texas at Dallas, Richardson, Texas 75080, USA}

\author{Aron W. Cummings}
\affiliation{Catalan Institute of Nanoscience and Nanotechnology (ICN2), CSIC and BIST, Campus UAB, Bellaterra, 08193 Barcelona, Spain}

\author{Alexander Cerjan}
\affiliation{Center for Integrated Nanotechnologies, Sandia National Laboratories, Albuquerque, New Mexico 87185, USA}

\author{Wei Pan}
\affiliation{Sandia National Laboratories, Livermore, California 94551, USA}

\author{Fan Zhang}
\affiliation{Department of Physics, University of Texas at Dallas, Richardson, Texas 75080, USA}

\author{Catalin D. Spataru}
\affiliation{Sandia National Laboratories, Livermore, California 94551, USA}

\begin{abstract}
Nano-patterned semiconductor interfaces offer a versatile platform for creating quantum metamaterials and exploring novel electronic phenomena. In this study, we illustrate this concept using artificial graphene---a metamaterial featuring distinctive properties including Dirac and saddle points. We demonstrate that introducing additional nano-patterning can open a Dirac band gap, giving rise to what we term artificial hexagonal boron nitride (AhBN). The calculated valley Chern number of AhBN indicates the presence of topological valley Hall states confined to Dirac-gap domain walls. A key question is whether these one-dimensional edge states are topologically protected against disorder, given their vulnerability to Anderson localization. To this end, we perform band structure and electronic transport simulations under experimentally relevant disorder, including charge puddles and geometric imperfections. Our results reveal the resilience of the domain wall states against typical experimental disorder, particularly while the AhBN band gap remains open. The localization length along the domain wall can reach several microns---several times longer than the bulk electron mean free path---even though the number of bulk transport channels is greater. To enhance the effectiveness of the low-dissipation domain wall channel, we propose ribbon geometries with a large length-to-width ratio. These findings underscore both the potential and challenges of AhBN for low-energy, power-efficient microelectronic applications. 
\end{abstract}
\maketitle

\section{Introduction}
Topological insulators (TIs)~\cite{hasanColloquium2010,moorebirth2010} are materials characterized by an insulating bulk but conducting surface, edge, or corner states. These dissipationless boundary modes arise from the contrast between the nontrivial topological invariant(s) of a TI's electronic structure and the trivial topology of states like the vacuum. This promise of efficient energy transfer mechanisms has generated significant interest in understanding how topological defects influence transport channels~\cite{teoTopological2010}. One such defect occurs when a two-dimensional (2D) material hosts a one-dimensional (1D) domain wall, characterized by a sharp discontinuity in a topological invariant across the interface. A key example of such an invariant is the valley Chern number~\cite{doi:10.1073/pnas.1308853110,martinTopological2008,zhangSpontaneous2011a,yaoEdge2009,doi:10.1021/nl201941f}, a quantity associated with Berry curvature integrated over a single valley of an electronic band structure. The resulting valley Hall effect that arises from the difference of valley Chern numbers across a topological domain wall has been observed not only in bilayer graphene and transition-metal dichalcogenide electronic devices~\cite{aldenStrain2013,juTopological2015,huangHightemperature2024,zhuImaging2022,dengEpitaxially2025}, but also in acoustic and photonic crystals~\cite{Lu2017,Gao2018,Yang2020,zhangBrought2018}. As the effect seems to be universal to both quantum and classical systems, it offers a promising avenue for practical energy or information transfer applications.

The performance of valley Hall systems based on traditional solid-state materials can, however, be challenged by the presence of disorder~\cite{doi:10.1073/pnas.1308853110}, as has been recently demonstrated in the context of photonic topological interfaces~\cite{Rosiek2023}. One approach to address this issue involves implementing novel, tunable materials. In this context, we propose an innovative mechanism for realizing valleytronics on the surface of an industry-grade semiconductor using an advanced fabrication technology.

First, we consider artificial graphene (AG) as our primary quantum metamaterial. AG can be constructed by nano-engineering a 2D electron gas (2DEG) in a semiconductor heterostructure (such as AlSb/InAs/AlSb), and introducing a triangular antidot lattice via interferometric lithography~\cite{panLithographyDefined2025}. The antidots repel electrons into regions of low potential, resulting in an effective honeycomb lattice where the pseudo-carbon atoms are located. Because changing the nano-pattern design is  relatively straightforward, AG offers greater versatility than natural graphene. As a critical advantage, the AG lattice constant and the vertical depth of the patterned antidots may be strategically chosen. Then, we introduce secondary antidots in the pattern to open a bulk band gap in AG. As this mimics the structure of hexagonal boron nitride (hBN), we dub this metamaterial `artificial hexagonal Boron Nitride' (AhBN). Opening a gap in this manner in a traditional graphene-like system reveals valley Chern numbers~\cite{doi:10.1073/pnas.1308853110,martinTopological2008,zhangSpontaneous2011a,yaoEdge2009,doi:10.1021/nl201941f}. Therefore, we aim to understand the role of band topology and the effects of disorder on our AhBN system.

To investigate these effects, we first consider the introduction of disorder in the form of a fluctuating potential induced by hypothetical charge puddles~\cite{dassarmaElectronic2011}. As a semiconductor heterostructure contains nearby charge impurities or similar defects, analyzing the effect of this type of disorder is important for understanding how the valley Hall effect may survive under realistic conditions. Then, we consider a second type of disorder resulting from the geometrical variation in the nano-patterned antidots in the 2DEG that produce the AG/AhBN. Both types of disorder are relatively short-range for AG/AhBN, meaning that their relevant disorder length scales are shorter than the inter-pseudoatomic distance $L$; therefore, valley-Hall topological protection may be less effective, as it may not completely suppress inter-valley scattering~\cite{doi:10.1073/pnas.1308853110}. We analyze the impact of these types of disorder as a function of the disorder strength, on both the AhBN Dirac band gap and the Dirac-gap domain wall state. 

Because the valley Chern number is not strictly quantized and the domain wall states are 1D in nature, they are susceptible to Anderson localization. We aim to understand the interplay between disorder-induced localization and the valley Hall effect, which provides some level of protection against localization. In particular, we seek to quantify the localization length of the domain wall state to accurately determine whether our system can demonstrate the valley Hall effect over length scales relevant to nano patterned devices. For realistic disorder strengths, we find that the localization length spans several microns, {\it i.e.}\ an order of magnitude larger than the AG periodicity $L$ and significantly larger than the typical mean free path ($<1$ $\upmu$m) of the underlying 2DEG. Therefore, the proposed meta-graphene valley Hall states, combined with the wide tunability and ultra-high quality of semiconducting superlattices, provide a promising platform for future applications in electronic devices with low-power dissipation. 

\begin{figure}
    \centering
    \includegraphics[width=\columnwidth]{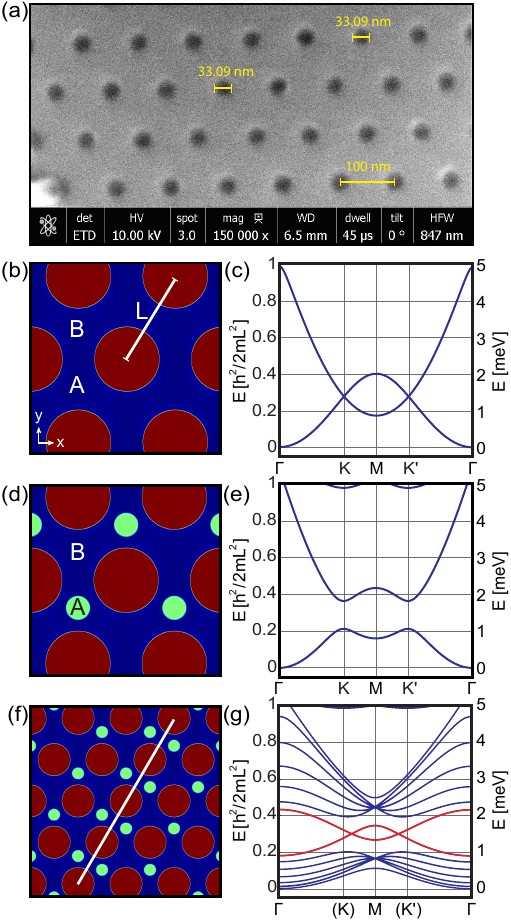}
    \caption{(a) SEM image of the experimentally produced AG with a lattice constant of $L=100$ nm. (b) Potential profile of AG. The red holes (antidots) indicate potential barriers of strength $V_1$ that low-energy electrons cannot penetrate, resulting in the graphene-like band structure in (c). (d) Addition of holes of strength $V_2$ at the A sites of the lattice breaks inversion symmetry (and sublattice symmetry) and opens the band gap at the K/K' points, as indicated in the band structure in (e). (f) Potential profile of AhBN with a domain wall. The white line indicates the domain wall, which acts as a mirror line and contains an inversion center of the entire system. (g) The band structure projected along the domain wall direction that has a translational symmetry of AG. Each red band indicates a pair of domain wall states counter-propagating at the K/K' valleys; there are two parallel domain walls in our setup to ensure the periodic boundary condition.}
    \label{fig:ab_ag}
    \label{fig:abba_ag}
\end{figure}

\section{Model}
\subsection{Artificial Graphene and Artificial hBN}
The model for artificial graphene (AG) that we employ is the continuum model described by Park and Louie~\cite{Park2009}. Fig.~\ref{fig:ab_ag}(a) shows a typical Scanning electron microscope (SEM) image of AG with a lattice constant of $L = 100$ nm in an InAs quantum well. We use a triangular lattice of `muffin tin' to represent the antidots as potential barriers that repel electrons. Along with the kinetic term describing an electron gas characterized by an effective mass $m$, we express the total Hamiltonian in a plane wave basis. A similar approach, utilizing a tight-binding basis, has recently been applied---and validated experimentally---to describe AG structures~\cite{spataruTopological2025,panLithographyDefined2025}. 
In our calculations, we set the potential strength of the primary triangular `muffin-tin' lattice to $V_1=10 \frac{h^2}{2mL^2}$. 
This nano patterning results in the formation of AG, with the pseudo-atoms arranged in a conjugate honeycomb lattice, as indicated by the sites labeled A and B in Fig.~\ref{fig:ab_ag}(b). In reciprocal space, this produces a graphene-like band structure with Dirac points at the K and K' points and saddle points at the three M points, as shown in Fig.~\ref{fig:ab_ag}(c). The diameter of the antidots in our model is chosen to be $D_1 = \frac{L}{2}$, where $L = 100$~nm is the AG lattice constant. The antidots are expected to alter mostly the electronic behavior of 2DEG electrons with energies on the order of $V_1$, with the higher-energy electrons less affected. A secondary muffin-tin lattice with a potential strength $V_2$ is added at the A sites, breaking the inversion symmetry (and sublattice symmetry) between the A and B sites. Using $V_2 = 0.8 \frac{h^2}{2mL^2}$ and diameter $D_2 = \frac{L}{4}$ opens a band gap of about $0.15 \frac{h^2}{2mL^2}$ at the K/K' Dirac points~\cite{RevModPhys.82.1959}. We term this gapped AG system `artificial hexagonal Boron Nitride' (AhBN). Plots of the real-space potential profiles for AG and AhBN along with their low-energy bands are shown in Figs.~\ref{fig:ab_ag}(b,d) and Figs.~\ref{fig:ab_ag}(c,e), respectively.

\begin{figure}
    \centering
    \includegraphics[scale=1]{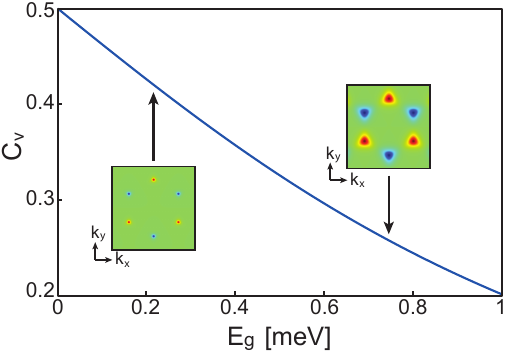}
    \caption{Plot of valley Chern number $C_v$ versus energy band gap $E_g$. Insets: momentum-space distribution of the Berry curvature at indicated values of $E_g$. The regions exhibiting nonzero curvature correspond to the K/K' points. For $V_1=10\frac{h^2}{2mL^2}$, $V_2 = 0.8\frac{h^2}{2mL^2}$, $D_1=\frac{L}{2}$, $D_2=\frac{L}{4}$, and $L=100$~nm, we obtain $E_g = 0.76$~meV and $C_v =\frac{1}{4}$.}
    \label{fig:berry}
\end{figure}
    
Breaking the inversion symmetry in graphene produces Berry curvature localized near the K/K' points. Because of time-reversal symmetry, the Berry curvatures at the two valleys are equal in magnitude but opposite in sign~\cite{doi:10.1073/pnas.1308853110,RevModPhys.82.1959}. A similar phenomenon occurs in AG; in the insets of Fig.~\ref{fig:berry} we illustrate how the Berry curvature seeps out of the K and K' points as the gap opens in AhBN. Additionally, we explicitly calculate the valley Chern numbers~\cite{fukuiChern2005}.
When the strength of the secondary potential approaches our chosen $V_2$, the valley Chern numbers reach approximately $C_v = \pm\frac{1}{4}$. This represents a significant deviation from the ideal case of $C_v =\pm \frac{1}{2}$~\cite{doi:10.1073/pnas.1308853110,RevModPhys.82.1959}, indicating that our chosen parameters test the limits of the valley Hall effect.

Leveraging existing knowledge of topological transport and bulk-edge correspondence~\cite{doi:10.1073/pnas.1308853110,PhysRevLett.71.3697}, we juxtapose two regions of AhBN against each other, as indicated in Fig.~\ref{fig:abba_ag}(f). Due to the nonzero difference between the valley Chern numbers on the two sides of the domain wall, this configuration is expected to produce topological edge states. Figure~\ref{fig:abba_ag}(g) shows the calculated band structure projected to the domain wall direction: there is a pair of in-gap states, counter-propagating at the two valleys and confined to its host domain wall. Note that, because of periodic boundary conditions, the supercell accommodates two parallel domain walls, resulting in two red bands of domain wall states in Fig.~\ref{fig:abba_ag}(g).

\section{Impact of Disorder}
\subsection{Charge Puddle Disorder}\label{subsec:charge_puddle_disorder}
To establish how robust these domain wall states are against disorder~\cite{doi:10.1021/nl201941f,doi:10.1073/pnas.1308853110,liTopological2011}, 
we first consider disorder whose length scale is much smaller than the lattice constant $L$, since $L$ in our AG is on the mesoscopic scale. Adhering to the type of disorder observed in graphene experiments, we first analyze the effect of charge-puddle disorder~\cite{Zhang2009}. Note that while the charge puddles are smooth across the graphene lattice, they exhibit sub-lattice-scale variation in AG---a key distinction we emphasize.

\begin{figure}
    \centering
    \includegraphics[scale=1]{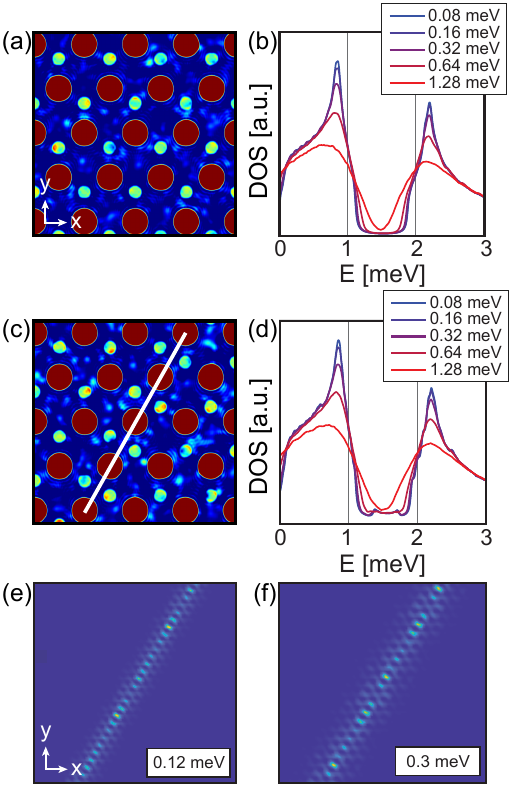}
    \caption{(a) Potential profile of AhBN with charge puddle disorder. 
    (b) Calculated DOS for (a) at different disorder strengths.  The bulk band gap closes at a critical disorder strength. 
    (c) The same as (a) but with a domain wall. (d) Calculated DOS for (c) at different disorder strengths. 
    The DOS within the bulk gap remains constant, indicating that the domain wall states survive on average until the disorder is sufficient to close the bulk gap. 
    (e,f) Local DOS of the domain wall states at the midgap energy ($1.5$~meV) for two charge-puddle configurations.}
    \label{fig:radii}
    \label{fig:puddles}
\end{figure}

We assume that the individual charge puddles take the form of Gaussians, with a standard deviation of $\sigma = 5$ nm for each charge puddle, and that there are $100$ charge puddles per AG unit cell on average (this corresponds to a relatively dirty sample with $1.1\times 10^{12}$~puddles/cm$^2$). The total disorder strength $U$ is defined as the standard deviation of a normal distribution from which the strength of the potential is sampled. 
Our calculations are performed using a $12 \times 12$ supercell for bulk AG and AhBN, and a $12 \times 24$ supercell for AhBN with two domain walls---corresponding to two $12 \times 12$ domains---ensuring periodic boundary conditions.
We consider more than $150$ random realizations of charge puddles, and two representative configurations are shown in Figs.~\ref{fig:puddles}(a,c). 

As shown in the density of states (DOS) in Fig.~\ref{fig:puddles}(b), the bulk band gap of AhBN remains open until the disorder strength $U$ approaches the magnitude of the gap. Fig.~\ref{fig:puddles}(d) shows the DOS of the AhBN system with two domains of opposite Dirac gaps. Evidently, a constant DOS persists within the gap until it is closed by disorder, indicating the presence of robust topological domain wall states. 
In Figs.~\ref{fig:puddles}(e,f) we present the local DOS of typical domain wall states at the midgap energy ($1.5$~meV). We note that for both charge puddle disorders with strengths $0.12$ and $0.3$~meV, the domain wall states appear to be extended. 
Additionally, it is clear that the localization of these states increases with the disorder strength. 
We will provide a more quantitative assessment using transport simulations below.

\subsection{Geometric Disorder}
\begin{figure}
    \centering
    \includegraphics[scale=1]{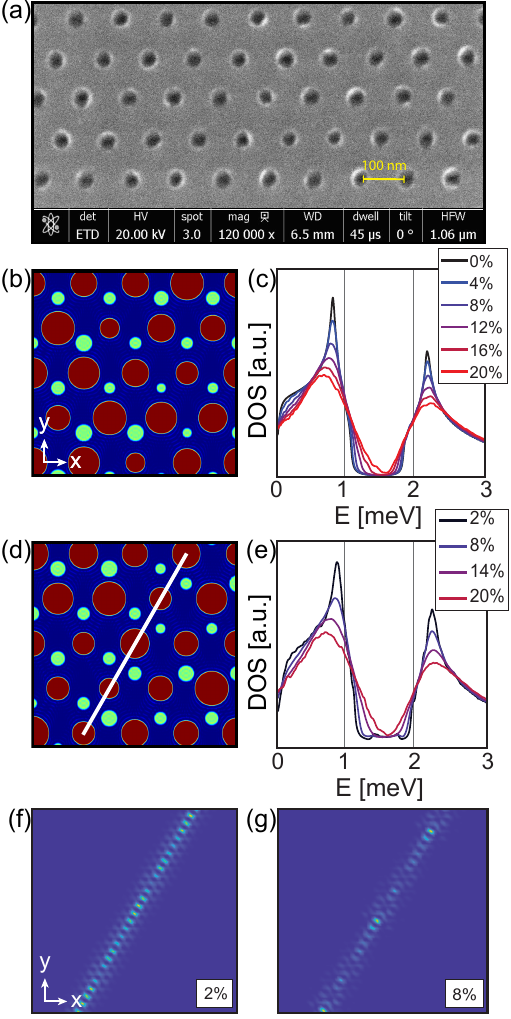}
    \caption{(a) SEM image of an AG sample exhibiting geometric disorder with a strength of approximately $9\%$. (b) Potential profile of AhBN with geometric disorder. (c) Calculated DOS for (b) at different disorder strengths, given in maximum percentage deviation of the antidot radius from the perfect case. 
    (d) The same as (b) but with a domain wall.  (e) Calculated DOS for (d) at different disorder strengths. 
    (f,g) Local DOS of the domain wall states at the midgap energy ($1.5$~meV) for two geometric disorder configurations.}
    \label{fig:geom}
\end{figure}

In addition to charge puddle disorder, we also account for random variations in the radii of individual antidots in our model, as this geometric aberration is relevant to the experimental construction of the AhBN system, as exemplified in Fig.~\ref{fig:geom}(a). We characterize the disorder strength by the maximum permitted deviation of the antidot radius from its unperturbed value, expressed as a percentage. The random radii are sampled from a uniform distribution within this range. Our DOS calculations for geometric disorder are performed using a $12\times12$ supercell for bulk AhBN, and a $12\times24$ supercell for AhBN with domain walls, utilizing approximately $200$ random configurations. 

Figs.~\ref{fig:geom}(b,d) illustrate how both the bulk and domain wall real-space potentials may change under geometrical disorder for a particular 
random configuration. Fig.~\ref{fig:geom}(c) shows the calculated average DOS, indicating that the band gap does not close until there is a deviation of $20\%$ from the perfect antidot lattice, which is significantly greater than the typically observed experimental deviation ($\leq5\%$). We further observe from Fig.~\ref{fig:geom}(e) that the finite DOS of the domain wall states signifies their presence within the gap, persisting up to the point at which the bulk gap closes. Figures~\ref{fig:geom}(f,g) show the extended domain wall states at the midgap energy ($1.5$~meV) for geometric deviations of $2\%$ and $8\%$. 
Similar to what is observed for charge puddle disorder, the domain wall states localize over several unit cells, and the localization increases with the deviation. Our DOS results allow us to refer to transport calculations for a more complete characterization of the localization of domain wall states.

\subsection{Reflection Invariant and Topological Protection}

\begin{figure}
\includegraphics[width=\columnwidth]{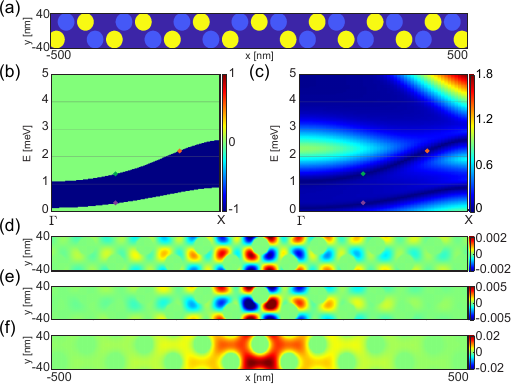}
\caption{(a) Geometry of a reflection symmetric AhBN ribbon with a domain wall. 
(b)-(c) Energy-resolved marker $\zeta_E^{\mathcal{R}}$ (b) and local gap $\mu_{E}$ (c) 
as a function of energy $E$ and wavevector $k_y$ along the domain wall direction. 
(d)-(f) Wavefunctions $\textrm{Re}[\psi_{(k_y,E)}(x)]$ of the topological domain wall-localized states 
at $E = 1.37$~meV (d), $E = 2.21$~meV (e), and $E = 0.32$~meV (f), which are marked by the green, orange, and purple dots in (b)-(c), respectively.}
\label{fig:sl1}
\end{figure}

In addition to AhBN exhibiting the valley Chern number, a reflection-symmetric domain wall is also associated with a topological invariant rendered by this symmetry. This invariant protects the existence of the interface-localized conduction channel against disorder that preserves the reflection symmetry. Furthermore, it also provides a measure of the protection---associated with the existence of the domain wall state---even in the presence of disorder that breaks this symmetry.

To classify the reflection-symmetric AhBN structure, we use the spectral localizer framework \cite{Loring2015,Loring2017,Loring2020,CerjanAPL}, which provides both a suite of local topological markers and an associated measure of topological protection. In particular, a reflection-symmetric 2D structure can be classified using the symmetry-reduced spectral localizer \cite{Cerjan2024crystalline}
\begin{equation}
\tilde{L}_{E}^{\mathcal{R}}(X,H) = (H-E\mathbf{1} + i\kappa X)\mathcal{R}.
\end{equation}
Here, $X$ is the position operator in the direction perpendicular to the reflection symmetry $\mathcal{R}$, $E$ is a choice of energy typically taken to be the Fermi energy, $\kappa$ is a scaling parameter on the order of the bulk band gap divided by the length of the system in the $x$-direction, $\kappa \sim E_{\textrm{gap}}/L$, and $\mathbf{1}$ is the identity. As $X\mathcal{R} = -\mathcal{R}X$ and $H\mathcal{R} = \mathcal{R}H$, $\tilde{L}_{E}^{\mathcal{R}}$ is Hermitian, and as such it possesses an energy-resolved integer invariant of matrix homotopy,
\begin{equation}
\zeta_E^{\mathcal{R}}(X,H) = \frac{1}{2}\textrm{sig}[\tilde{L}_{E}^{\mathcal{R}}(X,H)] \in \mathbb{Z}. \label{eq:zeta}
\end{equation}
Here, $\textrm{sig}[M]$ denotes the signature of a Hermitian matrix $M$, defined as its number of positive eigenvalues minus its number of negative eigenvalues. As eigenvalues of Hermitian matrices must move continuously, $\textrm{sig}[M]$ cannot change without one of its eigenvalues first becoming zero, i.e., without $M$ becoming non-invertible. Thus, the eigenvalue of $\tilde{L}_{E}^{\mathcal{R}}$ closest to zero carries information about a system's topological protection against reflection-symmetry preserving perturbations. Specifically, the system's local gap can be defined as
\begin{equation}
\mu_{E}(X,H) = \min(|\textrm{spec}[\tilde{L}_{E}^{\mathcal{R}}(X,H)]|),
\end{equation}
which by Weyl's inequality \cite{weyl_asymptotische_1912,Bhatia1997} requires that a change in the system's topology is only possible if a symmetry-preserving perturbation $\delta H$ is strong enough to close the local gap, $\Vert \delta H \Vert > \mu_{(x,E)}$. Here, $\textrm{spec}[M]$ denotes the spectrum of $M$. Finally, bulk-boundary correspondence manifests in the spectral localizer framework because closings of the local gap, where $\mu_{E} \approx 0$, guarantee the existence of nearby states \cite{cerjan_quadratic_2022}. Altogether, Eq.~\eqref{eq:zeta} distinguishes reflection-symmetric systems based on whether their associated spectral localizers can be path-continued to one another while remaining invertible and preserving $\mathcal{R}$. 

To demonstrate the topology associated with the domain wall in AhBN, we consider a reflection-symmetric ribbon of AhBN 
that we simulate using the approach in ~\cite{spataruTopological2025}, assuming $L=80$ nm. We consider open boundary conditions along the $x$-direction and periodic boundary conditions along the $y$-direction as shown in Fig.~\ref{fig:sl1}(a). Thus, we can use $\zeta_E^{\mathcal{R}}(X,H(k_y))$ and $\mu_{E}(X,H(k_y))$ to classify the system's topology and associated protection, choosing $\mathcal{R}$ to be the mirror symmetry along the domain wall. In doing so, we numerically observe that the change in the topological index and gap closing correspond to the domain wall-localized state, as seen in  Figs.~\ref{fig:sl1}(b-f). 

Of course, experimentally relevant fabrication disorder is not reflection symmetric. Nevertheless, the in-gap domain wall-localized states still exhibit some lingering protection stemming from Weyl's inequality. To demonstrate this remnant of protection, we consider the full spectral localizer for a 2D system,
\begin{multline}
L_{(x,y,E)}(X,Y,H) =  \\
\left[ \begin{array}{cc}
    H - E \mathbf{1} & \kappa(X-x\mathbf{1}) - i\kappa(Y-y\mathbf{1}) \\
    \kappa(X-x\mathbf{1}) + i\kappa(Y-y\mathbf{1}) & -(H - E \mathbf{1}) 
    \end{array} \right],
\end{multline}
and corresponding local gap
\begin{equation}
\mu_{(x,y,E)}(X,Y,H) = \min(|\textrm{spec}[L_{(x,y,E)}(X,Y,H)]|).
\end{equation}
Here, $X$ and $Y$ are the position operators of the 2D system, and $(x,y,E)$ is a choice of location in position-energy space where the spectral localizer and local gap are being evaluated. The key point is that locations where $\mu_{(x,y,E)} \approx 0$ still guarantee the existence of nearby states, regardless of whether the system exhibits any non-trivial topological index. Moreover, the rate at which $\mu_{(x,y,E)}$ can change in response to perturbations to the system is still bound by Weyl's inequality \cite{CerjanAPL}
\begin{equation}
    \mu_{(x,y,E)}(X,Y,H + \delta H) \le \mu_{(x,y,E)}(X,Y,H) + \Vert \delta H \Vert,
\end{equation}
where now $\delta H$ is an arbitrary perturbation to the system's Hamiltonian that need not preserve any symmetries. 

\begin{figure}[t!]
\includegraphics[width=\columnwidth]{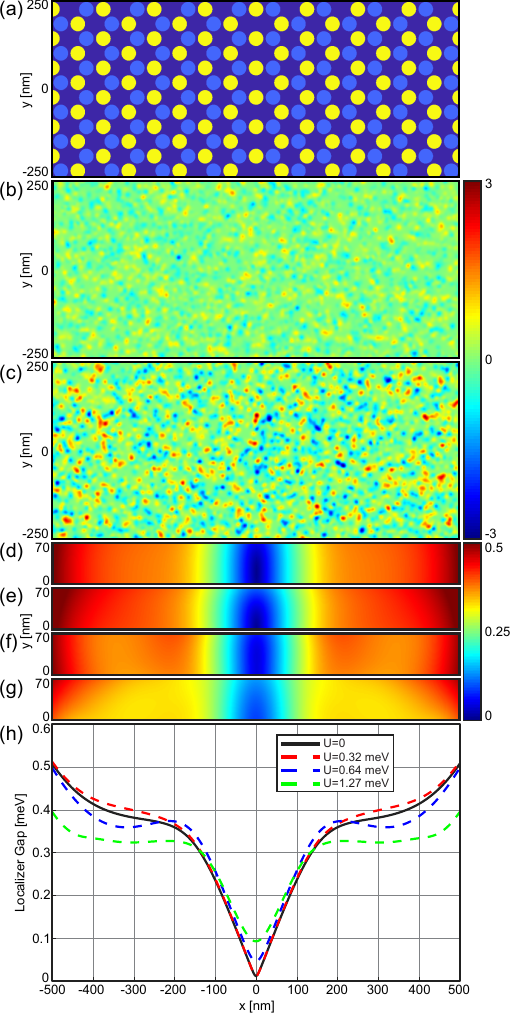}
\caption{(a) Geometry of a finite AhBN flake with a reflection-symmetric domain wall in the absence of disorder. 
(b)-(c) Distribution of the added charge-puddle disorder with standard deviation $\sigma = 5$ nm, $10^{12}$ charge puddles/cm$^2$, 
and disorder strength $U=0.32$~meV (b) and $U=0.64$~meV (c).
(d)-(g) Local gap $\mu_{(x,y,E)}$ at fixed energy for $U=0$ (d), $U=0.32$~meV (e), $U=0.64$~meV (f), and $U=1.27$~meV (g). 
(h) Line cuts of $\mu_{(x,y,E)}$ for fixed $y=0$ and $E=1.5$~meV, for values of $U$ from (d)-(g). The contribution to $\mu_{(0,y,E)}$ from disorder is negligible until $U$ approaches $E_{\textrm{gap}}$. }
\label{fig:sl2}
\end{figure}

The robustness of the domain wall-localized states in AhBN can be confirmed using a large-but-finite 2D AhBN lattice, shown in Fig.~\ref{fig:sl2}(a). Here, the local gap of the clean system approaches zero along the domain wall at energies where the topological index changes, Fig.~\ref{fig:sl2}(d). 

We next consider a disorder potential induced by charge puddles. As above, each individual charge puddle is modeled as a Gaussian with a standard deviation of $\sigma = 5$ nm, and the concentration of charge puddles is $10^{12}$/cm$^2$. The total disorder strength $U$ is defined as the standard deviation of a normal distribution from which the potential strength is sampled.
Upon addition of disorder, examples of which are shown in Figs.~\ref{fig:sl2}(b,c), the near closing of the local gap is seen to persist, Figs.~\ref{fig:sl2}(e-h), demonstrating the robustness of the domain wall-localized states against arbitrary disorder, though without providing any guarantees on the conductivity that these states may provide. Note, as the symmetry protecting the topology is broken by the disorder, $\zeta_E^{\mathcal{R}}(X,H+\delta H)$ is no longer well-defined as when $[\delta H,\mathcal{R}]\ne0$, $\tilde{L}_{E}^{\mathcal{R}}(X,H+\delta H)$ is not Hermitian. A quantitative assessment of their conductive properties requires transport simulations.

\subsection{Transport Simulations}
We now examine disorder on the scale of the AG lattice constant $L$, and its impact on transport through the domain wall state. We treat the disorder by first reducing the AhBN system to a single-orbital tight-binding representation on a honeycomb lattice, with the Hamiltonian
\begin{equation}
\hat{H} = \sum\limits_i \varepsilon^\x{A/B}_i \hat{c}^\dagger_i \hat{c}_i + \sum\limits_{\left<ij\right>} t_{ij} \hat{c}^\dagger_i \hat{c}_j,
+ \sum\limits_i \varepsilon^\x{dis}_i \hat{c}^\dagger_i \hat{c}_i,
\end{equation}
where $\hat{c}^\dagger_i$ ($\hat{c}_i$) is the creation (annihilation) operator of an electron at honeycomb lattice site $i$. The band gap of AhBN is captured by the onsite potential $\varepsilon^\x{A/B}_i = \pm \Delta / 2$ at lattice sites A and B respectively, while $t_{ij}$ is the hopping between the nearest neighbor sites $i$ and $j$. To capture lattice-scale disorder, we include the third term and employ an Anderson model for $\varepsilon^\x{dis}_i$~\cite{DJThouless_1970, PhysRev.109.1492}, where the potential of each lattice site is shifted by a random value. Here the random disorder potential is chosen according to a normal distribution centered around the mid-gap (the zero-energy reference) with standard deviation $U$. To model the AhBN lattice in Fig.~\ref{fig:ab_ag}, we set $\Delta = 0.78$~meV, $t_{ij} = 0.85$~meV, and $L = 100$~nm.

\begin{figure}
\includegraphics[width=\columnwidth]{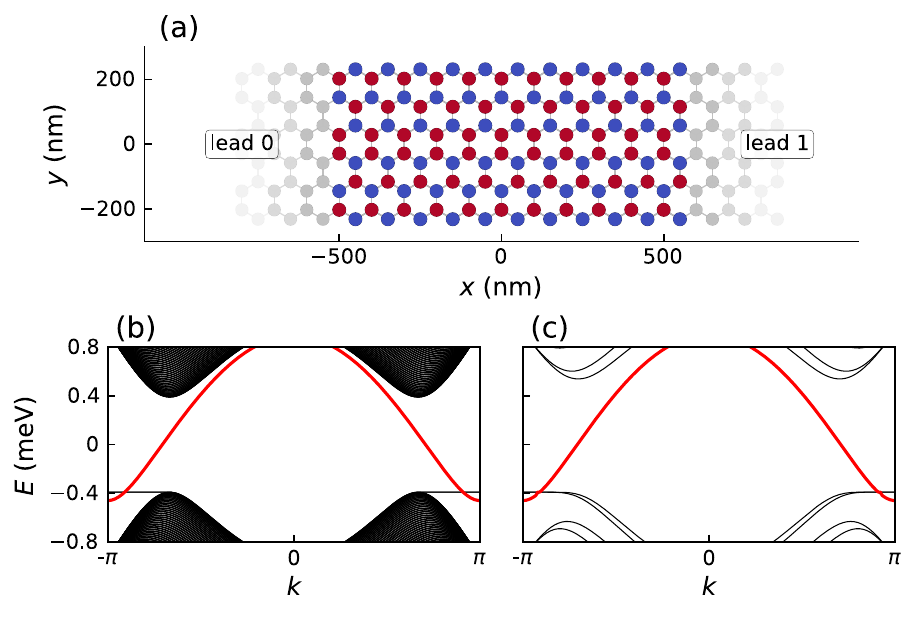}
\caption{(a) Setup for the transport simulations, which calculate transmission from lead 0 to lead 1. Red/blue lattice sites correspond to an onsite energy of $\pm\Delta/2$. (b) Band structure of the system in panel (a) for a channel width of 20~$\upmu$m. Black bands indicate the bulk bands of AhBN, while the red band indicates the domain wall state localized around $y=0$. (c) The same as (b) but for a channel width of $1$~$\upmu$m.}
\label{fig:tr_setup}
\end{figure}

To calculate transport through the domain wall state, we use the setup depicted in Fig.~\ref{fig:tr_setup}(a), built using the \textit{pybinding} package~\cite{pybinding}. We consider a finite channel of AhBN, with the color indicating the staggered onsite potential at each lattice site. For the top half of the channel we set $\varepsilon^\x{A/B}_i = \pm \Delta / 2$ and the opposite sign for the bottom half, thereby forming a domain wall at $y = 0$. For the purpose of visualization, only a relatively narrow channel is shown in Fig.~\ref{fig:tr_setup}(a). In our transport simulations we consider channels with widths of $W = 200 L = 20$~$\upmu$m 
and $W = 10 L = 1$~$\upmu$m, while varying the channel length $\lch$ up to $\lch = 10$~$\upmu$m.

We calculate the transmission from the left to the right of the channel, from lead 0 to lead 1, using the Landauer-Büttiker formalism implemented in the \textit{kwant} simulation package~\cite{kwant}. For a given disorder strength $U$, we calculate the energy- and channel length-dependent conductance $G(E,\lch) = 2e^2/h \cdot T(E,\lch)$, where $T$ is the transmission, $e$ is the electron charge, and $h$ is Planck's constant~\cite{Datta_1995}.
For each channel length we consider 200 random disorder configurations, from which we calculate the typical conductance $\gtyp = \exp\left( \left< \ln(G) \right> \right)$, where $\left< ... \right>$ indicates the disorder average~\cite{Anderson1980, Beenakker1997}. We then extract the energy- and channel length-dependent resistance as $R(E,\lch) = 1 / \gtyp(E,\lch)$. In all the cases we consider relatively weak disorder, $U = \Delta/6 = 0.13$ meV, which is expected to preserve the band gap and the domain wall state.

\subsubsection{Bulk Limit}

\begin{figure}
\includegraphics[width=\columnwidth]{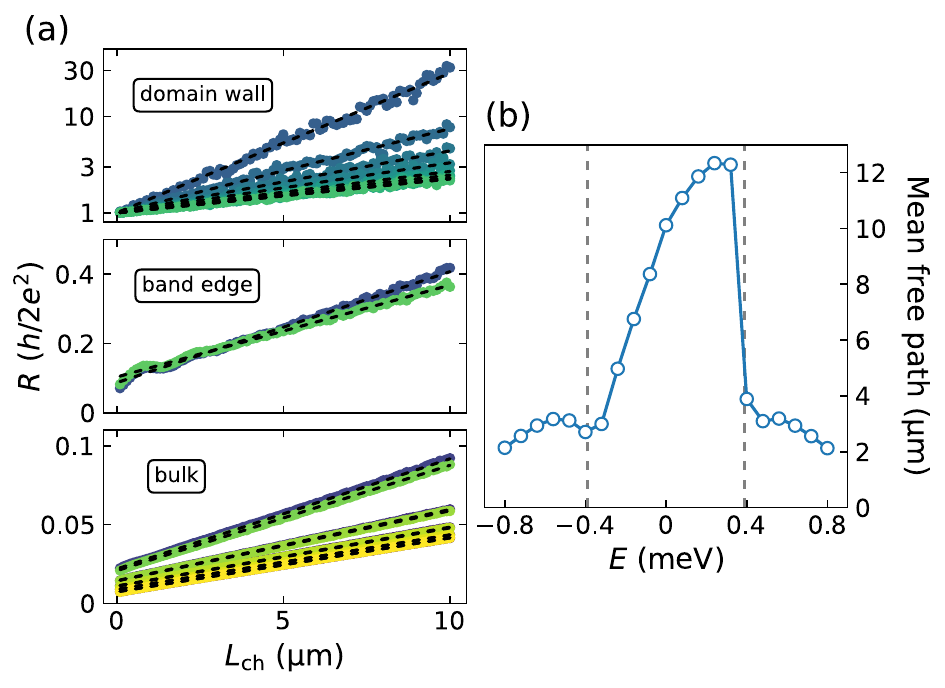}
\caption{(a) Resistance of AhBN vs. channel length for a wide channel ($W=20$~$\upmu$m) with a disorder strength of $U = \Delta/6 = 0.13$~meV. The top panel is for states within the bulk band gap ($|E| < 0.39$~meV), corresponding to the domain wall state. The middle panel is for states near the band edges ($|E| = 0.39$~meV), and the bottom panel is for bulk states ($|E| > 0.39$~meV). Colored symbols are numerical data, and black dashed lines are fits to extract the mean free path. (b) Mean free paths extracted from (a), indicating enhanced transport in the domain wall state compared to the bulk states. The vertical dashed lines mark the band edges.}
\label{fig:tr_weak}
\end{figure}

First we consider a very wide system, with a channel width $W = 200L = 20$ $\upmu$m, to mimic the bulk limit of AhBN. Fig.~\ref{fig:tr_setup}(b) shows the band structure of this setup in the absence of disorder. Similar to Fig.~\ref{fig:abba_ag}(f), the bulk bands have a gap of $\Delta = 0.78$~meV, and the domain wall state (shown in red) traverses the gap. The flat bands at $E \approx -0.39$~meV correspond to the zigzag edge states on the top and bottom edges of the channel. 
In Fig.~\ref{fig:tr_weak}(a) we show the length-dependent resistance over a range of energies ($|E| < 0.8$~meV). As indicated by the labels, the highest resistance corresponds to energies associated with the midgap domain wall state ($|E| < 0.39$~meV), the next group of curves corresponds to energies near the band edge ($|E| = 0.39$~meV), and the lowest-resistance curves correspond to the bulk states in the conduction and valence bands ($|E| > 0.39$~meV).

At short channel lengths, the resistance is determined by the number of modes available for transport. This is why the domain wall state has the highest resistance --- within the band gap there is only one mode. To better compare the transport properties across all energies, we use the scaling of $R$ vs. $\lch$ to extract the mean free path. In the bulk and at the band edges, the resistance increases linearly with channel length, as seen in the middle and bottom panels of Fig.~\ref{fig:tr_weak}(a). In this ohmic regime of transport, the mean free path can be extracted by fitting the simulation data to~\cite{Datta_1995}
\begin{equation}
\label{eq:mfp_ohm}
R(E,\lch) = \rc(E) \left( 1 + \frac{\lch}{\lmfp(E)} \right),
\end{equation}
where $\rc$ is the contact resistance and $\lmfp$ is the mean free path. The black dashed curves in Fig.~\ref{fig:tr_weak}(a) indicate the fits to the numerical data. 

Meanwhile, within the band gap the resistance of the domain wall state increases exponentially with channel length, as indicated in the top panel of Fig.~\ref{fig:tr_weak}(a). This is consistent with transport in a 1D channel, where the exponential scaling may be directly connected to the mean free path~\cite{Thouless1973, Beenakker1997}. For the case of one propagating mode, the mean free path can be extracted from~\cite{Beenakker1997}
\begin{equation}
\label{eq:mfp_loc}
R(E,\lch) = \frac{h}{2e^2} \exp(\lch /\lmfp(E)).
\end{equation}

We plot the energy-dependent mean free path in Fig.~\ref{fig:tr_weak}(b). Within the bulk, the mean free path falls within 2-3~$\upmu$m. Meanwhile, it reaches more than 12~$\upmu$m within the band gap, indicating the relative robustness of the domain wall state against weak disorder. However, despite the longer mean free path at the interface, the overall resistance remains much lower in the bulk AhBN region due to the larger number of modes available for transport.

\subsubsection{AhBN Ribbon}
In the previous subsection we have revealed that while the mean free path of the topological interface state is much higher than that of the AhBN bulk states, the overall resistance is also much higher for the interface state because of the single mode nature. This then raises the question: is there a situation where the midgap domain wall state shows comparable or better transport properties compared to the bulk of AhBN? In principle, this may be achieved in a much narrower system, as lateral confinement should reduce the number of bulk conduction modes and could also degrade the bulk transport properties. We examine this here by considering the same setup as before but with a channel width of $10$ unit cells, $W = 10L = 1$~$\upmu$m.

\begin{figure}
\includegraphics[width=\columnwidth]{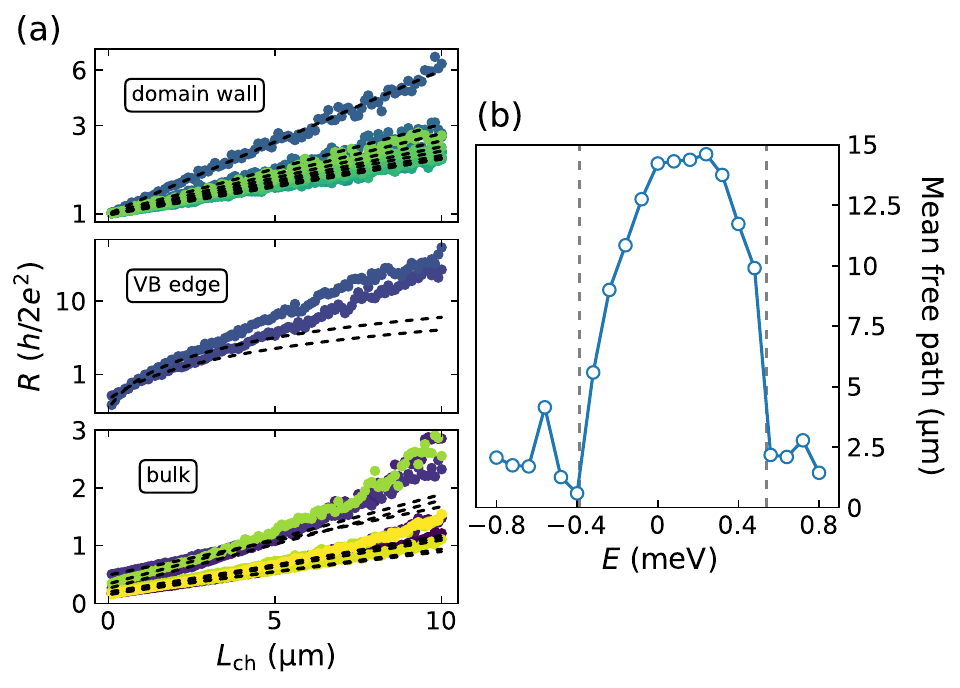}
\caption{(a) Resistance of AhBN vs.\ channel length for a narrow channel ($W = 1$ $\upmu$m) with a disorder strength of $U = \Delta/6 = 0.13$~meV. The top panel is for states within the bulk band gap ($-0.39 < E < 0.54$~meV), corresponding to the domain wall state. The middle panel is for states near the valence band edge ($E = -0.4$ and $-0.48$~meV), and the bottom panel is for bulk states ($E < -0.48$ and $E > 0.54$~meV). Colored symbols are numerical data, and black dashed lines are fits to extract the mean free path. (b) Mean free path extracted from (a), indicating enhanced transport in the domain wall state compared to the bulk states. The vertical dashed lines mark the band edges.}
\label{fig:tr_weak_narrow}
\end{figure}

Figure~\ref{fig:tr_setup}(c) shows the band structure of this narrower system, which is  similar to that of the wider system shown in Fig.~\ref{fig:tr_setup}(b), but with a few key changes. First, the number of subbands in the bulk is significantly reduced, which will lead to a significant increase in the bulk resistance. Second, owing to the lateral confinement, the size of the band gap has increased a bit, to the range $-0.39 < E < 0.54$~meV.  
Figure~\ref{fig:tr_weak_narrow} shows the length-dependent resistance and the mean free path of this narrower system. The domain wall states exhibit localized transport similar to the case of the wider system. By contrast, the states in the bulk and at the band edge, which are ohmic in the case of the wider system, now exhibit a metal-to-insulator transition at a channel length of a few microns. 
In Fig.~\ref{fig:tr_weak_narrow}(b) we plot the mean free path, extracted using the same procedure as before. 
For the bulk and band-edge states we fit Eq.~\eqref{eq:mfp_ohm} to the resistance for $\lch < 1$~$\upmu$m, for which transport is still in the ohmic regime~\footnote{We find the same mean free path by fitting to Eq.~\eqref{eq:mfp_loc} in the localized regime for $\lch > 5$ $\upmu$m and scaling by the number of modes $N$, $\lmfp \to \lmfp \cdot 2/(N+1)$ \cite{Beenakker1997}}. Evidently, the mean free paths of the domain wall state and the bulk states remain similar to the wider channel case, with the former slightly increased and the latter slightly reduced.

Owing to the metal-to-insulator transition, as well as the reduced number of modes available for transport, 
the resistance of the bulk AhBN states is much higher in this narrower system than in the wider one, with the resistance being comparable to that in the bulk gap. Therefore, a narrower channel appears to be a system where a topological interface state can provide a transport advantage over the bulk AhBN.

\section{Conclusion}
We have demonstrated that nano-patterned meta-graphene can exhibit a designable band gap at the Dirac point, 
giving rise to an artificial analog of hexagonal boron nitride with nontrivial valley-projected band topology. 
Through an investigation of the DOS of the domain wall edge state and the behavior of both the spectral localizer and the local gap in the presence of two disorder mechanisms, 
we have sought to test the robustness of our domain wall state through transport simulations. 
Our numerical analysis shows that the domain wall state in AhBN remains robust even in the presence of experimentally relevant disorder, 
with localization lengths reaching beyond ten microns, extending substantially longer than the bulk mean free path. 
These results suggest that while topological transport mediated by the domain wall states is resilient, 
practical implementation is limited by residual bulk conduction. 
As such, we propose that high aspect ratio ribbon geometries can enhance the domain wall-dominated transport, 
offering a viable route toward integrating such systems in microelectronic applications. 
Our findings highlight the potential of nano-engineered 2D materials for exploring topological phenomena beyond natural crystals.

\section{acknowledgments}
The authors acknowledge support from the Laboratory Directed
Research and Development program at Sandia National
Laboratories. This work was performed, in part, at the
Center for Integrated Nanotechnologies, an Office of
Science User Facility operated for the U.S. Department
of Energy (DOE) Office of Science. Sandia National
Laboratories is a multimission laboratory managed and
operated by National Technology and Engineering
Solutions of Sandia, LLC, a wholly owned subsidiary
of Honeywell International, Inc., for the U.S. DOE’s
National Nuclear Security Administration under Contract
No. DE-NA-0003525.
ICN2 is funded by the CERCA Programme Generalitat de Catalunya and supported by the Severo Ochoa Centres of Excellence programme, Grant CEX2021-001214-S, funded
by MCIN/AEI/10.13039.501100011033.
P.P. and F.Z. (UT Dallas) were supported by the NSF under grants DMR-2414726, DMR-1945351, and DMR-2324033 
and by the Welch Foundation under grant AT-2264-20250403. 
P.P. and F.Z. acknowledge the Texas Advanced Computing Center (TACC) for providing supercomputing resources.

\bibliography{refs}

\end{document}